\documentclass{pasj01}
\Received{$\langle$reception date$\rangle$}
\Accepted{$\langle$acception date$\rangle$}
\Published{$\langle$publication date$\rangle$}

\begin{document}

\title{Bright Type II Supernova 2023ixf in M101: A Quick Analysis of the 
Early-Stage Spectra and Near-Infrared Light Curves}
\author{Masayuki Yamanaka$^{1}$, 
Mitsugu Fujii$^{2}$, and
Takahiro Nagayama$^{3}$}
\altaffiltext{1}{Amanogawa Galaxy Astronomy Research Center (AGARC), Graduate School of Science and Engineering, Kagoshima University, 1-21-35 Korimoto, Kagoshima, Kagoshima 890-0065, Japan}
\altaffiltext{2}{Fujii Kurosaki Observatory, 4500 Kurosaki, Tamashima, Kurashiki, Okayama 713-8126, Japan}
\altaffiltext{3}{Graduate School of Science and Engineering, Kagoshima University, 1-21-35 Korimoto, Kagoshima, Kagoshima
890-0065, Japan}
\email{yamanaka@sci.kagoshima-u.ac.jp}


\maketitle

\begin{abstract}
 We present early-stage analyses of low-resolution ($R=1000$) 
 optical spectra and near-infrared light curves of the bright 
 Type II supernova (SN II) 2023ixf 
 in the notable nearby face-on spiral galaxy M101, 
 which were obtained from $t=1.7$ to $8.0$ d. 
 Our first spectrum 
 showed remarkable emission 
 features of Balmer series, He~{\sc ii}, N~{\sc iii}, C~{\sc iv}, and N~{\sc iv} with a strong blue continuum.
 Compared with the SNe II showing flash-ionized features, we suggest that 
 this SN could be categorized into the high-luminosity SNe II with a nitrogen/helium-rich 
 circumstellar material (CSM), e.g., SNe 2014G, 2017ahn, and 2020pni. 
 The H~{$\alpha$} emission line can be tentatively explained by 
 a narrower component with a velocity of 
 $<300$ km~s$^{-1}$ and a broader one with $\sim2200$ km~s$^{-1}$.
 The near-infrared light curves were well consistent
 with those of the another luminous SN 2017ahn, and
 its absolute magnitudes locate on the bright end
 in the luminosity distribution of SNe II.
 These observational facts support that SN 2023ixf is well consistent {\bf with high-luminosity SNe II showing evidences of a dense nitrogen/helium-rich CSM.} 
\end{abstract}

\section{Introduction}

 {\bf The discovery} of a supernova (SN) explosion in an extremely nearby 
galaxy provides a lot of understandings of astrophysics. 
It makes us to obtain the spectroscopic and multi -band data
soon after explosion. Multi-wavelength 
all-sky survey and follow-up observations may give a strong 
constraint on the physical properties of an explosive event.
{\bf The wide-field high-cadence surveys have} developed rapidly, 
and transients including SNe has been discovered at 
their very young stages, while some highly-active amateur astronomers 
also exceedingly contribute to the discovery and prompt follow-up 
of transients.

 Recently, prompt spectroscopic observations of
{\bf some infant Type II SNe were carried out soon after their discovery.}
The spectra are often dominated by 
highly-excited emission features with a blue continuum 
\citep{Gal-Yam2014,Yaron2017}. The emission features
support evidence of the presence of gas surrounding 
the progenitor \citep{Gal-Yam2014}. 
{\bf \citet{Terreran2022} illustrated that high-excitation lines of 
hydrogen, carbon and helium were present in the early spectra 
of the Type II SN 2020pni.}
They discussed the possibility of a high nitrogen/helium abundance in its circumstellar medium (CSM).

 SN 2023ixf was discovered at 14.9 magnitude in
the outskirt of the prominent face-on spiral galaxy M101 
at 17:27:15.000 (UT) on May 19 in 2023 by Koichi Itagaki 
\citep{Itagaki2023ixf}. {\bf After that, prompt multi-wavelength 
and -messenger observations and quick analyses were carried 
out all over the world \citep{Thwaites2023,Kawai2023,Maund2023,YZhang2023,Grefenstette2023,Perley2023}.}
Thereafter, this SN brightened steadily but 
rapidly and reached a maximum brightness of $V\sim10.8$ 
mag around May 24 \citep{Fowler2023}. 
{\bf The detection of progenitor activity in pre-explosion data was also 
reported
\citep{Szali2023}.}
SN 2023ixf is the second brightest SN among all-type SNe discovered since SN 1987A in the optical bands, following Type Ia SN 2011fe discovered in M101 which showed a maximum brightness at $10.0$ mag in the $V$-band \citep{Pereira2013,KZhang2016}.
\footnote{Many efforts are being made in "bright supernova" page 
to discover bright supernovae and follow up on their brightness in https://rochesterastronomy.org/} 

 In this Letter, we report the early-phase 
 low-resolution spectroscopic observations and
 the near-infrared imaging observations. In \S 2,
 we show our observations and 
 data reduction. In \S 3, we show the 
 early-stage spectral and near-infrared light
 curve evolution. We also show 
 the line identification and the 
 comparison with other SNe II. Finally,
 we present our discussion and 
 conclusion in \S 4.

\begin{figure}
 \begin{center}
  \includegraphics[width=80mm]{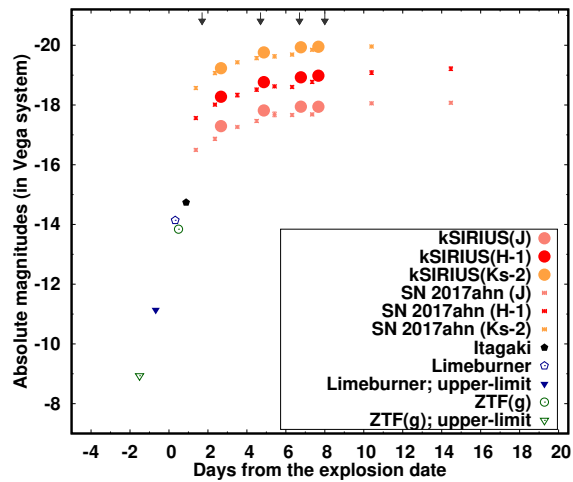}
 \end{center}
 \caption{The multi-band absolute magnitude light curves of SN 2023ixf. Filled open circles denote $JHKs$-band photometric data 
 obtained by kSIRIUS. 
 The asterisk-shape symbols denote $JHKs$-band light curves 
 of SN 2017ahn \citep{Tartagrlia2021}. 
{\bf The downward arrows correspond to the epochs of our spectroscopic data.}
 The black 
 filled-pentagon symbol denotes the discovery magnitude, and 
 the black filled-downward triangle denotes the upper-limit 
 magnitude \citep{Itagaki2023ixf}.
 The blue open pentagon symbol denotes the pre-discovery 
 magnitude, and the downward blue open triangle denotes
 the upper-limit magnitude by Stephen Limeburner \citep{Limeburner2023}. 
 {\bf The green open circle denotes the pre-discovery $g'$-band magnitude and
 the downward green triangle denotes the upper-limit magnitude by \citet{Perley2023b}.}}
\end{figure}

\begin{figure*}
 \begin{center}
  \includegraphics[width=140mm]{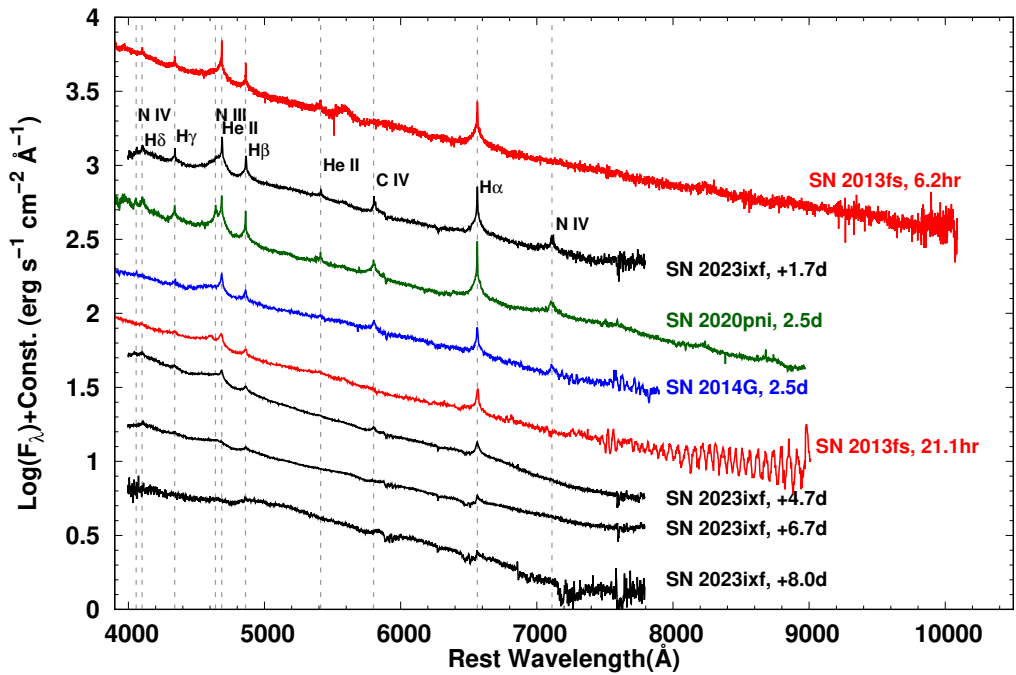}
 \end{center}
 \caption{Spectra of SN 2023ixf obtained at $t=1.7$, $4.7$, $6.7$, and $8.0$d, compared with two spectra of SNe 2013fs \citep{Yaron2017}, 2014G \citep{Terreran2016}, and 2020pni \citep{Terreran2022}. Emission lines 
 of Balmer series, He~{\sc ii}, C~{\sc iii}, N~{\sc iii}, C~{\sc iv}, and N{\sc iv} were
 identified through the comparison with SNe 2014G and 2020pni. }
\end{figure*}

\section{Observations and Data reduction}
 The spectroscopic observations were carried out using
the spectrograph installed to the 0.4-m reflector on four
nights from May 20.5 to May 26.7 UT at the 
Fujii Kurosaki Observatory (FKO) in Okayama.
The wavelength resolution was $R=1000$ and 
the coverage was 4000-7800 \AA. The data reduction
was performed according to the standard manner.
{\bf The wavelength calibration was performed
using telluric lines}, and the flux calibration was done 
using the spectra of the hot-temperature standard stars
observed on the same night as SN 2023ixf. 
The telluric absorption in the spectra were also corrected.
 The near-infrared imaging observations were
performed using kSIRIUS, the near-infrared simultaneous 
$JHKs$-band camera \footnote{kSIRIUS was developed in 
the same concept to SIRIUS (the near-infrared
simultaneous three-band camera; \citep{Nagayama2003} but the current array size is smaller (320x256). The size of the field-of-view is $3.7'\times2.9'$
and the pixel scale is $0.69"$ pix$^{-1}$.} 
attached to the Cassegrain focus of the 1.0-m telescope from
May 20.7 to May 26.5 on four nights at the Iriki 
Observatory in Kagoshima. The data reduction
was performed according to the standard manner to
the NIR imaging data. 
{\bf Magnitudes were obtained through the 
point-spread-function (PSF) technique 
using standard IRAF tasks such as DAOPHOT \citep{Stetson1987} .}

\section{Results}
 \subsection{Light curve properties}

 Figure 1 shows our near-infrared multi-band absolute magnitude
light curves. The discovery magnitude by Koichi 
Itagaki \citep{Itagaki2023ixf}
is also plotted in this figure. 
The pre-discovery detection and the upper-limit magnitudes reported by \citet{Limeburner2023} and \citet{Perley2023b}.
{\bf A number of pre-discovery and upper-limit magnitudes} were 
collected and the explosion date was estimated to be 
MJD 60082.83 \citep{Yaron2023}. 
In this Letter, we define this epoch as $t=0$.

The distance to the M101 is $6.4$Mpc which was 
well calibrated 
using SN Ia 2011fe \citep{Matheson2012} 
and Cepheid stars \citep{Macri2001,Shappee2011}. 
The corresponding distance modulus was $\mu=29.0$ mag.
{\bf We measured an equivalent width (EW) of $\sim1.0$\AA\ for the Na~{\sc i}D doublet in our spectrum at $t=1.7$ d. The host extinction was then computed using the empirical relation obtained in \citet{Poznanski2012} 
corresponding to $A_v\simeq0.64\,\rm{mag}$, although the relation
has a uncertainty if EW$>0.6$ \AA.}
The host galactic extinction was corrected based on this assumption.
{\bf After the correction, we confirmed that 
the color were almost similar to those of 
another Type II SN 2017ahn \citep{Tartagrlia2021}.}

{\bf The $JHKs$-band light curves  were compared with those of SN 2017ahn.}
The light curves of SN 2023ixf impressively follow those of 
SN 2017ahn. It may make us predict the light curve evolution of SN 2023ixf.
The absolute magnitudes were calculated to be 
$M_{J}=-17.9$, $M_{H}=-17.9$, and $M_{Ks}=-17.9$
mag on at $t=7.8$ d, respectively. 
{\bf According to this comparison the near-infrared light curves of SN 2023ixf may still rise to a peak magnitude similar to SN 2017ahn ($M_{H}\sim-18.0$ mag).}
These locate around the bright end in the luminosity 
distribution of SNe II \citep{Anderson2014}.

\subsection{Line identification and spectral evolution}

 {\bf Our first spectrum showed strong emission lines} with 
a blue continuum (see Figure 2). The spectrum was compared 
with those of other SNe II showing narrow emission lines 
with a blue continuum. The emission lines of 
H~$\alpha$, H~$\beta$, H~$\gamma$ and H~$\delta$ were identified 
in our spectra by comparison.
The H~$\alpha$ emission line has multiple components (see Figure 3).  
The feature can be tentatively resolved by 
two Gaussian function with full-width at half-maximum (FWHM)
of 300 and 2200 km~s$^{-1}$, although adding 
the third component 
to the broad one may improve fit accuracy. The width of the 
narrower component was comparable {\bf to} the resolution.

  The highly-excited nitrogen (N~{\sc iv} $\lambda$7111) 
 and carbon (C~{\sc iv}$\lambda$ 5801) emission lines were also 
 identified by a comparison with SNe 2014G and 2020pni.
 These emission lines were slightly separated 
 into two components. The N~{\sc iv} $\lambda$4058 was also marginally detected. 
 These high-excitation level emission lines 
 showed only the narrow component in our data.

 The emission feature at 4640-4680 \AA\ was 
 found and contributed by He~{\sc ii}$\lambda$4686, Bowen N~{\sc iii}$\lambda\lambda$4634,4642, and C~{\sc iii}$\lambda$4642. 
 The He~{\sc ii}$\lambda$4686 feature consists of the multiple components including 
 the narrow and the broad components. 
 The He~{\sc ii}$\lambda$5412 showed only the narrow component. 

  At $t=4.7$ d, the intensity of some emission lines 
 quite weakened. The narrow emission lines of Balmer
 series greatly attenuated and the broad component cannot 
 almost be recognized. Similarly, the mutliplet 
 of He~{\sc ii}, C~{\sc iii}, and N~{\sc iii}
 also became weak. The C~{\sc iv}$\lambda$5801 feature was
 marginally detected by a comparison with 
 the earlier phase one. The N~{\sc iv} $\lambda$7111
 emission line was invisible. The continuum was still unambiguous. 

 At $t=6.7$ d, the intensity of the H~{$\alpha$} 
became very weak and the narrow P Cyg profile developed  (see Figure 4). 
The structure of the absorption component was complicated. The blue end 
of the absorption reached -5000 km~s$^{-1}$.
The H~{$\beta$} emission line was still seen. The H~{$\gamma$} was not 
significantly detected. At $t=8.0$ d, 
the overall features became weak while the 
H{$\alpha$} marginally showed the P Cyg profile.

\begin{figure}
 \begin{center}
  \includegraphics[width=80mm]{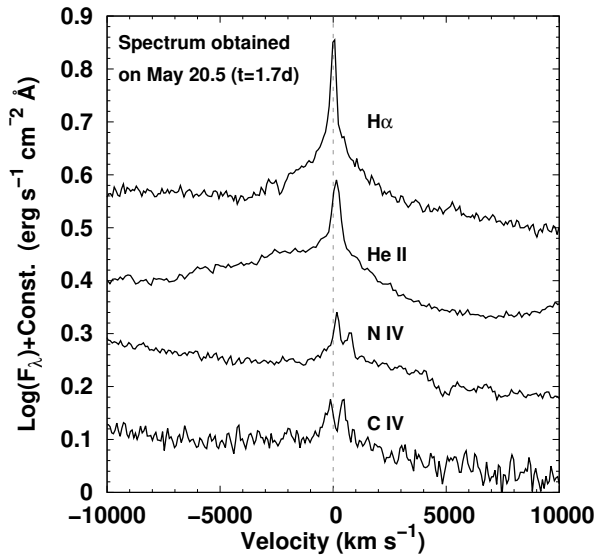}
 \end{center}
 \caption{Close up the emission line profile of H~{$\alpha$}, 
 He~{\sc ii}, N~{\sc iv}$\lambda$7111, and 
 C~{\sc iv}$\lambda$5801. The H$\alpha$ emission line can be explained 
 by narrow ($\sim300$ km~s$^{-1}$) and broad ($\sim$2200 km~s$^{-1}$) components.}
\end{figure}

 \subsection{Comparison with other SNe II}
 Figure 2 shows our first spectrum at $t=1.7$ d, which was 
compared with SNe 2013fs \citep{Yaron2017}, 2014G 
\citep{Terreran2016}, and 2020pni \citep{Terreran2022} 
\footnote{These spectral data were downloaded in WISERep (http://wiserep.weizmann.ac.il/)}.
Spectra of comparison objects exhibit flash ionized features
with a blue continuum. The emission lines are 
Balmer series, He~{\sc ii}$\lambda4686$, 
C~{\sc iv}$\lambda$5801, and N~{\sc iv}$\lambda$7111. 
Broad components such as H$\alpha$ and He~{\sc ii}
were also seen as in other objects.

 Overall line profile in our spectrum had a excellent 
match with those of SNe 2014G, and 2020pni.
{\bf These all exhibited narrow H~{\sc i} and He~{\sc ii}, as well as 
prominent C~{\sc iv} and N~{\sc iv} emission lines.}
SN 2017ahn also has 
high-excitation N~{\sc iv} emission lines \citep{Tartagrlia2021}. 
As we will discuss in \S 4, these
objects form {\bf a remarkable subclass of SNe IIP.}
 
 Compared with SN 2013fs, no signature of O~{\sc v} 
emission line were detected in SN 2023ixf during 
our observations. On the
other hand, the emission features of N~{\sc iv} and 
C~{\sc iv} were invisible in SN 2013fs at t=$6.2$ and $21.1$ 
hours after its explosion. The H{$\alpha$} emission 
line evolution was similar among these objects.

\begin{figure}
 \begin{center}
  \includegraphics[width=80mm]{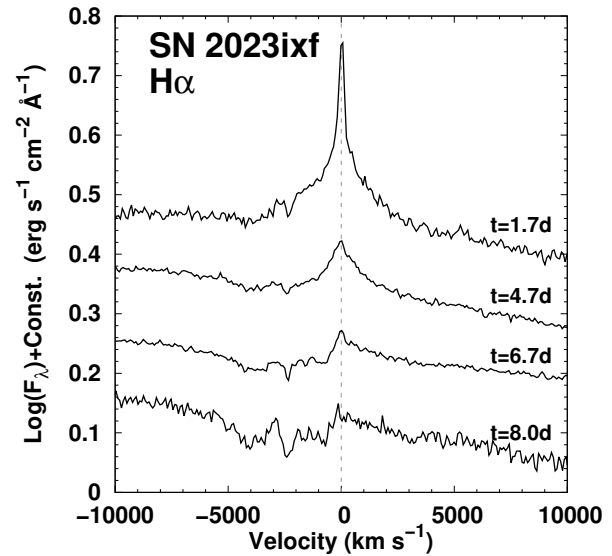}
 \end{center}
 \caption{The temporal evolution of the line profile of
 H{$\alpha$} from $t=1.7$ to $8.0$ d. The first 
 spectrum showed the multiple components with broad and 
 narrow strong emission lines. At $t=4.7$ d, 
 the narrow component attenuated. The feature 
 weakened as it evolved to the later epoch.}
\end{figure}

\section{Discussion and Conclusion}

 We presented the observational similarity of SN 2023ixf 
 with SNe 
 2014G, 2017ahn, and 2020pni based on our original
 early-stage spectral and near-infrared light curve data. 
 We showed that the near-infrared light curves and their 
peak absolute magnitudes of SN 2023ixf were 
well similar to those of SN 2017ahn.
 \citet{Terreran2022} discussed that the light curves and 
 the spectral evolution of SN 2020pni are 
well similar to those of SNe 2014G \citep{Terreran2016}, 
and 2017ahn \citep{Tartagrlia2021}.
The detection of the highly-excited nitrogen and carbon 
emission lines are also consistent with those of 
SNe 2014G and 2020pni. These facts support that 
SN 2023ixf could be categorized into this subclass.
 
\citet{Terreran2022} discussed the physical properties
of the CSM through the analysis of their flash-ionized features of SN 2020pni using the radiative transfer code, CMFGEN \citep{Hillier1998}. 
They suggested that the surrounding gas has a nitrogen/helium-rich abundance. 
Assuming that the high luminosity correlates with the large progenitor mass,
the high-mass progenitor may be a condition to induce the ejection 
of the nitrogen/helium-rich gas. 
However, \citet{Pledger2023} carried out the quick analysis 
of the HST data and presented that the progenitor mass 
would be $\sim12M_{\odot}$ assuming the single-star evolution .
The estimate of the radioactive nickel mass from the 
cobalt decay line would provide the firm progenitor mass 
with a small uncertainty.

\citet{Terreran2022} discussed that the large ejection 
of the envelope may occur within a few years just before explosion
from the analysis of the radio and optical spectral data.
They discussed that the interaction of the progenitor 
with a binary companion may trigger the ejection, 
but pointed out that the large fraction of the 
flash-ionized event \citep{Khazov2016,Bruch2022} disfavors 
the frequent interaction. For SN 2023ixf, the deep radio 
observations will be performed 
in the future. In order to unveil the nature of the 
CSM and the complex evolution of at the pre-explosion 
stage, the multi -mode and -wavelength analyses will be encouraged.

\begin{ack}
 We are grateful to graduate and undergraduate students for performing 
 the near-infrared observations. We thank Osamu Ohshima for the helpful advice. 
 This work was supported by Grant-in-Aid for Scientific Research (C) 
 22K03676. The Kagoshima University 1 m telescope is a member of the Optical and Infrared Synergetic Telescopes for Education and Research (OISTER) program funded by the MEXT of Japan.
\end{ack}







\bibliographystyle{apj}
\bibliography{addsample}

\begin{thebibliography}{}
\expandafter\ifx\csname natexlab\endcsname\relax\def\natexlab#1{#1}\fi

\bibitem[{{Anderson} {et~al.}(2014){Anderson}, {Gonz{\'a}lez-Gait{\'a}n},
  {Hamuy}, {Guti{\'e}rrez}, {Stritzinger}, {Olivares E.}, {Phillips},
  {Schulze}, {Antezana}, {Bolt}, {Campillay}, {Castell{\'o}n}, {Contreras}, {de
  Jaeger}, {Folatelli}, {F{\"o}rster}, {Freedman}, {Gonz{\'a}lez}, {Hsiao},
  {Krzemi{\'n}ski}, {Krisciunas}, {Maza}, {McCarthy}, {Morrell}, {Persson},
  {Roth}, {Salgado}, {Suntzeff}, \& {Thomas-Osip}}]{Anderson2014}
{Anderson}, J.~P., {Gonz{\'a}lez-Gait{\'a}n}, S., {Hamuy}, M., {et~al.} 2014,
  \apj, 786, 67

\bibitem[{{Bruch} {et~al.}(2022){Bruch}, {Gal-Yam}, {Yaron}, {Chen},
  {Strotjohann}, {Irani}, {Zimmerman}, {Schulze}, {Yang}, {Kim}, {Bulla},
  {Sollerman}, {Rigault}, {Ofek}, {Soumagnac}, {Masci}, {Fremling}, {Perley},
  {Nordin}, {Cenko}, {Ho}, {Adams}, {Adreoni}, {Bellm}, {Blagorodnova},
  {Burdge}, {De}, {Dekany}, {Dhawan}, {Drake}, {Duev}, {Graham}, {Graham},
  {Jencson}, {Karamehmetoglu}, {Kasliwal Shrinivas Kulkarni}, {Miller},
  {Neill}, {Prince}, {Riddle}, {Rusholme}, {Sharma}, {Smith}, {Sravan},
  {Taggart}, {Walters}, \& {Yan}}]{Bruch2022}
{Bruch}, R.~J., {Gal-Yam}, A., {Yaron}, O., {et~al.} 2022, arXiv e-prints,
  arXiv:2212.03313

\bibitem[{{Fowler} {et~al.}(2023){Fowler}, {Sienkiewicz}, \&
  {Dussault}}]{Fowler2023}
{Fowler}, M., {Sienkiewicz}, F., \& {Dussault}, M. 2023, Transient Name Server
  AstroNote, 143, 1

\bibitem[{{Gal-Yam} {et~al.}(2014){Gal-Yam}, {Arcavi}, {Ofek}, {Ben-Ami},
  {Cenko}, {Kasliwal}, {Cao}, {Yaron}, {Tal}, {Silverman}, {Horesh}, {De Cia},
  {Taddia}, {Sollerman}, {Perley}, {Vreeswijk}, {Kulkarni}, {Nugent},
  {Filippenko}, \& {Wheeler}}]{Gal-Yam2014}
{Gal-Yam}, A., {Arcavi}, I., {Ofek}, E.~O., {et~al.} 2014, \nat, 509, 471

\bibitem[{{Grefenstette}(2023)}]{Grefenstette2023}
{Grefenstette}, B. 2023, The Astronomer's Telegram, 16049, 1

\bibitem[{{Hillier} \& {Miller}(1998)}]{Hillier1998}
{Hillier}, D.~J., \& {Miller}, D.~L. 1998, \apj, 496, 407

\bibitem[{{Itagaki}(2023)}]{Itagaki2023ixf}
{Itagaki}, K. 2023, tNSTR, 1158, 1

\bibitem[{{Kawai} {et~al.}(2023){Kawai}, {Serino}, {Negoro}, {Mihara},
  {Nakajima}, {Kobayashi}, {Tanaka}, {Soejima}, {Kudo}, {Kawamuro}, {Yamada},
  {Tamagawa}, {Matsuoka}, {Sakamoto}, {Serino}, {Sugita}, {Hiramatsu},
  {Nishikawa}, {Yoshida}, {Tsuboi}, {Urabe}, {Nawa}, {Nemoto}, {Shidatsu},
  {Takahashi}, {Niwano}, {Sato}, {Higuchi}, {Yatsu}, {Nakahira}, {Ueno},
  {Tomida}, {Ishikawa}, {Ogawa}, {Kurihara}, {Ueda}, {Setoguchi}, {Yoshitake},
  {Nakatani}, {Yamauchi}, {Hagiwara}, {Umeki}, {Otsuki}, {Yamaoka}, {Kawakubo},
  {Sugizaki}, {Iwakiri}, \& {MAXI Team}}]{Kawai2023}
{Kawai}, N., {Serino}, M., {Negoro}, H., {et~al.} 2023, The Astronomer's
  Telegram, 16044, 1

\bibitem[{{Khazov} {et~al.}(2016){Khazov}, {Yaron}, {Gal-Yam}, {Manulis},
  {Rubin}, {Kulkarni}, {Arcavi}, {Kasliwal}, {Ofek}, {Cao}, {Perley},
  {Sollerman}, {Horesh}, {Sullivan}, {Filippenko}, {Nugent}, {Howell}, {Cenko},
  {Silverman}, {Ebeling}, {Taddia}, {Johansson}, {Laher}, {Surace},
  {Rebbapragada}, {Wozniak}, \& {Matheson}}]{Khazov2016}
{Khazov}, D., {Yaron}, O., {Gal-Yam}, A., {et~al.} 2016, \apj, 818, 3

\bibitem[{{Limeburner}(2023)}]{Limeburner2023}
{Limeburner}, S. 2023, Transient Name Server AstroNote, 128, 1

\bibitem[{{Macri} {et~al.}(2001){Macri}, {Calzetti}, {Freedman}, {Gibson},
  {Graham}, {Huchra}, {Hughes}, {Madore}, {Mould}, {Persson}, \&
  {Stetson}}]{Macri2001}
{Macri}, L.~M., {Calzetti}, D., {Freedman}, W.~L., {et~al.} 2001, \apj, 549,
  721

\bibitem[{{Matheson} {et~al.}(2012){Matheson}, {Joyce}, {Allen}, {Saha},
  {Silva}, {Wood-Vasey}, {Adams}, {Anderson}, {Beck}, {Bentz}, {Bershady},
  {Binkert}, {Butler}, {Camarata}, {Eigenbrot}, {Everett}, {Gallagher},
  {Garnavich}, {Glikman}, {Harbeck}, {Hargis}, {Herbst}, {Horch}, {Howell},
  {Jha}, {Kaczmarek}, {Knezek}, {Manne-Nicholas}, {Mathieu}, {Meixner},
  {Milliman}, {Power}, {Rajagopal}, {Reetz}, {Rhode}, {Schechtman-Rook},
  {Schwamb}, {Schweiker}, {Simmons}, {Simon}, {Summers}, {Young}, {Weyant},
  {Wilcots}, {Will}, \& {Williams}}]{Matheson2012}
{Matheson}, T., {Joyce}, R.~R., {Allen}, L.~E., {et~al.} 2012, \apj, 754, 19

\bibitem[{{Maund} {et~al.}(2023){Maund}, {Wiersema}, {Shrestha}, {Steele}, \&
  {Hume}}]{Maund2023}
{Maund}, J.~R., {Wiersema}, K., {Shrestha}, M., {Steele}, I., \& {Hume}, G.
  2023, Transient Name Server AstroNote, 135, 1

\bibitem[{{Nagayama} {et~al.}(2003){Nagayama}, {Nagashima}, {Nakajima},
  {Nagata}, {Sato}, {Nakaya}, {Yamamuro}, {Sugitani}, \&
  {Tamura}}]{Nagayama2003}
{Nagayama}, T., {Nagashima}, C., {Nakajima}, Y., {et~al.} 2003, Society of
  Photo-Optical Instrumentation Engineers (SPIE) Conference Series, Vol. 4841,
  {SIRUS: a near infrared simultaneous three-band camera}, ed. M.~{Iye} \&
  A.~F.~M. {Moorwood}, 459--464

\bibitem[{{Pereira} {et~al.}(2013){Pereira}, {Thomas}, {Aldering}, {Antilogus},
  {Baltay}, {Benitez-Herrera}, {Bongard}, {Buton}, {Canto}, {Cellier-Holzem},
  {Chen}, {Childress}, {Chotard}, {Copin}, {Fakhouri}, {Fink}, {Fouchez},
  {Gangler}, {Guy}, {Hillebrandt}, {Hsiao}, {Kerschhaggl}, {Kowalski},
  {Kromer}, {Nordin}, {Nugent}, {Paech}, {Pain}, {P{\'e}contal}, {Perlmutter},
  {Rabinowitz}, {Rigault}, {Runge}, {Saunders}, {Smadja}, {Tao},
  {Taubenberger}, {Tilquin}, \& {Wu}}]{Pereira2013}
{Pereira}, R., {Thomas}, R.~C., {Aldering}, G., {et~al.} 2013, \aap, 554, A27

\bibitem[{{Perley} {et~al.}(2023){Perley}, {Gal-Yam}, {Irani}, \&
  {Zimmerman}}]{Perley2023}
{Perley}, D.~A., {Gal-Yam}, A., {Irani}, I., \& {Zimmerman}, E. 2023, Transient
  Name Server AstroNote, 119, 1

\bibitem[{{Perley} \& {Irani}(2023)}]{Perley2023b}
{Perley}, D.~A., \& {Irani}, I. 2023, Transient Name Server AstroNote, 120, 1

\bibitem[{{Pledger} \& {Shara}(2023)}]{Pledger2023}
{Pledger}, J.~L., \& {Shara}, M.~M. 2023, arXiv e-prints, arXiv:2305.14447

\bibitem[{{Poznanski} {et~al.}(2012){Poznanski}, {Prochaska}, \&
  {Bloom}}]{Poznanski2012}
{Poznanski}, D., {Prochaska}, J.~X., \& {Bloom}, J.~S. 2012, \mnras, 426, 1465

\bibitem[{{Shappee} \& {Stanek}(2011)}]{Shappee2011}
{Shappee}, B.~J., \& {Stanek}, K.~Z. 2011, \apj, 733, 124

\bibitem[{{Stetson}(1987)}]{Stetson1987}
{Stetson}, P.~B. 1987, \pasp, 99, 191

\bibitem[{{Szalai} \& {Dyk}(2023)}]{Szali2023}
{Szalai}, T., \& {Dyk}, S.~V. 2023, The Astronomer's Telegram, 16042, 1

\bibitem[{{Tartaglia} {et~al.}(2021){Tartaglia}, {Sand}, {Groh}, {Valenti},
  {Wyatt}, {Bostroem}, {Brown}, {Yang}, {Burke}, {Chen}, {Davis},
  {F{\"o}rster}, {Galbany}, {Haislip}, {Hiramatsu}, {Hosseinzadeh}, {Howell},
  {Hsiao}, {Jha}, {Kouprianov}, {Kuncarayakti}, {Lyman}, {McCully}, {Phillips},
  {Rau}, {Reichart}, {Shahbandeh}, \& {Strader}}]{Tartagrlia2021}
{Tartaglia}, L., {Sand}, D.~J., {Groh}, J.~H., {et~al.} 2021, \apj, 907, 52

\bibitem[{{Terreran} {et~al.}(2016){Terreran}, {Jerkstrand}, {Benetti},
  {Smartt}, {Ochner}, {Tomasella}, {Howell}, {Morales-Garoffolo},
  {Harutyunyan}, {Kankare}, {Arcavi}, {Cappellaro}, {Elias-Rosa},
  {Hosseinzadeh}, {Kangas}, {Pastorello}, {Tartaglia}, {Turatto}, {Valenti},
  {Wiggins}, \& {Yuan}}]{Terreran2016}
{Terreran}, G., {Jerkstrand}, A., {Benetti}, S., {et~al.} 2016, \mnras, 462,
  137

\bibitem[{{Terreran} {et~al.}(2022){Terreran}, {Jacobson-Gal{\'a}n}, {Groh},
  {Margutti}, {Coppejans}, {Dimitriadis}, {Kilpatrick}, {Matthews}, {Siebert},
  {Angus}, {Brink}, {Filippenko}, {Foley}, {Jones}, {Tinyanont}, {Gall},
  {Pfister}, {Zenati}, {Ansari}, {Auchettl}, {El-Badry}, {Magnier}, \&
  {Zheng}}]{Terreran2022}
{Terreran}, G., {Jacobson-Gal{\'a}n}, W.~V., {Groh}, J.~H., {et~al.} 2022,
  \apj, 926, 20

\bibitem[{{Thwaites} {et~al.}(2023){Thwaites}, {Vandenbroucke}, {Santander}, \&
  {IceCube Collaboration}}]{Thwaites2023}
{Thwaites}, J., {Vandenbroucke}, J., {Santander}, M., \& {IceCube
  Collaboration}. 2023, The Astronomer's Telegram, 16043, 1

\bibitem[{{Yaron} {et~al.}(2023){Yaron}, {Bruch}, {Chen}, {Irani}, {Zimmerman},
  {Gal-Yam}, \& {Qin}}]{Yaron2023}
{Yaron}, O., {Bruch}, R., {Chen}, P., {et~al.} 2023, Transient Name Server
  AstroNote, 133, 1

\bibitem[{{Yaron} {et~al.}(2017){Yaron}, {Perley}, {Gal-Yam}, {Groh}, {Horesh},
  {Ofek}, {Kulkarni}, {Sollerman}, {Fransson}, {Rubin}, {Szabo}, {Sapir},
  {Taddia}, {Cenko}, {Valenti}, {Arcavi}, {Howell}, {Kasliwal}, {Vreeswijk},
  {Khazov}, {Fox}, {Cao}, {Gnat}, {Kelly}, {Nugent}, {Filippenko}, {Laher},
  {Wozniak}, {Lee}, {Rebbapragada}, {Maguire}, {Sullivan}, \&
  {Soumagnac}}]{Yaron2017}
{Yaron}, O., {Perley}, D.~A., {Gal-Yam}, A., {et~al.} 2017, Nature Physics, 13,
  510

\bibitem[{{Zhang} {et~al.}(2016){Zhang}, {Wang}, {Zhang}, {Zhang},
  {Ganeshalingam}, {Li}, {Filippenko}, {Zhao}, {Zheng}, {Bai}, {Chen}, {Chen},
  {Huang}, {Mo}, {Rui}, {Song}, {Sai}, {Li}, {Wang}, \& {Wu}}]{KZhang2016}
{Zhang}, K., {Wang}, X., {Zhang}, J., {et~al.} 2016, \apj, 820, 67

\bibitem[{{Zhang} {et~al.}(2023){Zhang}, {Fan}, {Zheng}, {Zhang}, \&
  {He}}]{YZhang2023}
{Zhang}, Y., {Fan}, Z., {Zheng}, J., {Zhang}, J., \& {He}, M. 2023, Transient
  Name Server AstroNote, 132, 1

\end{thebibliography}

\end{document}